\documentclass{iau} 
\usepackage{graphicx}

\title[On the circumstellar envelopes of SRs] 
{On the circumstellar envelopes of semi-regular long-period variables}

\author[J. J. D\'iaz-Luis et al.]   
{J. J. D\'iaz-Luis$^1$, J. Alcolea$^1$, V. Bujarrabal$^1$, M. Santander-Garc\'ia$^1$, M. G\'omez-Garrido$^2$
 \and J. -F. Desmurs$^1$}

\affiliation{$^1$Observatorio Astron\'omico Nacional (IGN), Spain \\ email: {\tt jjairo@oan.es} \\[\affilskip]
$^2$Instituto Geogr\'afico Nacional (IGN), Centro de Desarrollos Tecnol\'ogicos, Observatorio de Yebes, Spain}

\pubyear{2018}
\volume{343}  
\jname{Why Galaxies care about AGB stars}
\editors{F. Kerschbaum, M. Groenewegen, H. Olofsson, eds.}
\begin{document}

\maketitle

\begin{abstract}
The mass loss process along the AGB phase is crucial for the formation of circumstellar envelopes (CSEs), which in the post-AGB phase will evolve into planetary nebulae (PNe). There are still important issues that need to be further explored in this field; in particular, the formation of axially symmetric PNe from spherical CSEs. To address the problem, we have conducted high S/N IRAM 30m observations of $^{12}$CO \textit{J}=1-0 and \textit{J}=2-1, and $^{13}$CO \textit{J}=1-0 in a volume-limited unbiased sample of semi-regular variables (SRs). We also conducted Yebes 40m SiO \textit{J}=1-0 observations in 1/2 of the sample in order to complement our $^{12}$CO observations. We report a moderate correlation between mass loss rates and the $^{12}$CO(1-0)-to-$^{12}$CO(2-1) intensity ratio, introducing a possible new method for determining mass loss rates of SRs with short analysis time. We also find that for several stars the SiO profiles are very similar to the $^{12}$CO profiles, a totally unexpected result unless these are non-standard envelopes.
\end{abstract}

\firstsection 
\section{Introduction}

For low and intermediate mass stars ($\sim$1-8 M$_{\odot}$), the latest stages of their evolution are controlled by the mechanisms of mass loss that occur at these phases. In the AGB (and in general in the red giant or super-giant phases), these mass losses, with rates over 10$^{-5}$ M$_{\odot}$\,yr$^{-1}$, are crucial for the formation of circumstellar envelopes (CSEs), which in the post-AGB phase will evolve into planetary nebulae (PNe); see e.g. Olofsson (1999). There are still important issues that need to be further explained; in particular, the formation of axially symmetric PNe from spherical CSEs. CSEs around AGB stars are in general quite spherical, and expand isotropically at moderate speeds (10-25 km\,s$^{-1}$), at least at large scales (Neri et al. 1998; Castro-Carrizo et al. 2010). However, PNe display a large variety of shapes with high degree of symmetry; about 4/5 of the cases are far from showing spherically symmetric envelopes (Parker et al. 2006). This transformation from spherical CSEs to axially symmetric PNe is still a matter of debate since more than two decades (Balick $\&$ Frank 2002), though in the last years there is growing consensus that the presence of binary systems is the most likely explanation; see e.g. Jones $\&$ Boffin (2017). This would be supported by the large prevalence of multiple systems in solar-type stars in general, and in the AGB population in particular. Some studies reveal the presence of spiral patterns and bipolar outflows at smaller scales, providing observational support for the binary system scenario as the reason for such structures (e.g. Morris et al. 2006; Mauron $\&$ Huggins 2006; Maercker et al. 2012; Kim et al. 2015). Moreover, recent studies involve searching for UV excess emission, which is believed to be due to a main-sequence companion, from AGB stars using the GALEX archive (e.g. Sahai et al. 2008; Ortiz $\&$ Guerrero 2016). Almost 60$\%$ of the sample has a main-sequence companion of spectral type earlier than K0. It is believed that more than 50$\%$ of PNe harbour binary systems (Douchin et al. 2015) and almost 50$\%$ of solar-type stars in the solar neighborhood seem to form in binary or multiple systems (Raghavan et al. 2010). The main problem is that there are no complete and unbiased samples that resolve the mystery.

During the last years, the results from the increasing number of good interferometric maps of the CSEs of a particular type of evolved stars have attracted our attention. Semi-regular variables (SRs) are red giant stars showing quasi-regular or some irregular variations in the optical range (SRa and SRb variables, respectively), in contrast to the regular pulsators, the Mira-type variables (note that both SRs and Miras conform the Long-Period Variable stars, LPVs). In principle, the differences in their pulsation mode should not affect the envelope geometry and kinematics. However, a clear axial symmetry has been found in almost all CSEs around SRa and SRb variables that have been well mapped (using mm-wave interferometric observations of CO; see e.g. Castro-Carrizo et al. 2010; Hirano et al. 2014; Homan et al. 2017, 2018), in strong contrast to the CSEs of regular pulsators (Miras). In spite of a strong prevalence of non-spherical CSEs in SRs, we should take into account that the sample is limited and strongly biased. Hence, the question of whether SRs really have CSEs different from those of regular variables remains open.

\section{Observations}

Following these ideas, we conducted high S/N IRAM 30m observations of $^{12}$CO \textit{J}=1-0 and \textit{J}=2-1, and $^{13}$CO \textit{J}=1-0, as well as Yebes 40m SiO \textit{J}=1-0 observations, in a volume-limited unbiased sample of well characterized SRs. We selected all the variables in the General Catalog of Variable Stars (GCVS; Samus et al. 2012) with declinations above -25$^{\circ}$, and with information on the spectral/chemical type and variability period, for which the Hipparcos parallaxes (van Leeuwen 2007) are larger than 2 mas (all sources are closer than 500 pc; note that Gaia parallaxes were not available by the time we started the project). Moreover, we have further refined our sample by selecting only targets with IRAS 60 $\mu$m fluxes larger than 4 Jy (and with 12, 25, and 60 $\mu$m fluxes of quality 3; IRAS PSC 1988). This resulted in a list of 49 sources. We characterized the main properties of the CSEs, such as expansion velocity, mass loss rate, wind density, etc. 

\section{Preliminary results}

Lines were classified into four groups based on the $^{12}$CO line profile. Type 1 profiles are broad and symmetric lines with expansion velocities between 9 and 13 km\,s$^{-1}$. Type 2 profiles are narrow and symmetric lines with expansion velocities between 3 and 9 km\,s$^{-1}$. Type 3 profiles, however, are very strange line profiles, with very pronounced asymmetries and not compatible with standard optically thin lines already studied, with expansion velocities between 9 and 17 km\,s$^{-1}$. Finally, type 4 profiles are those lines which show two different components: one narrow component with low expansion velocity and another broad component with higher expansion velocity (like the addition of type 2+3 profiles).

\begin{figure}[h!]
 \centering
 \includegraphics[width=0.38\textwidth]{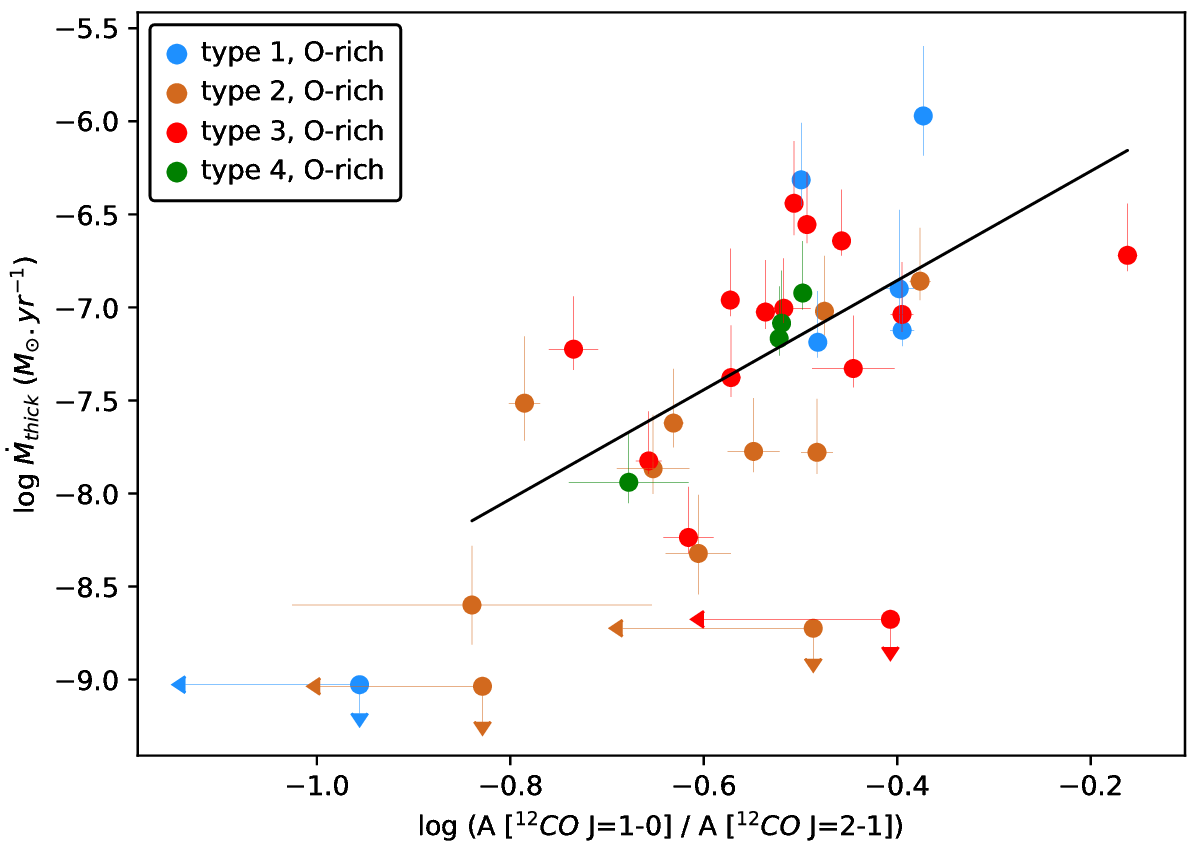} 
 \caption{Mass loss rates vs. $^{12}$CO(1-0)-to-$^{12}$CO(2-1) intensity ratio in oxygen-rich SRs. Note the moderate correlation between both parameters (r=0.66).}
   \label{fig1}
\end{figure}

We derived gas mass loss rates as discussed in Knapp $\&$ Morris (1985), by considering optically thick and optically thin envelopes. We also know that the envelope size is limited by the photodissociation of CO by the ambient interstellar radiation field and \textit{T$_{mb}$} as a function of the photodissociation radius is different for optically thick and optically thin cases. For optically thick cases, this variation depends a little on $\dot{M}$ and abundance of CO. The photodissociation radius has been estimated by using the latest results of Groenewegen (2017), which take into account the strength of the interstellar radiation field. Therefore, we calculated consistently mass loss rates and photodissociation radii by numerically solving the system in both cases.

We found a moderate correlation between mass loss rates and $^{12}$CO line intensity ratio for O-rich SRs (r=0.66; see Fig. 1). For each star the optically thick estimate of mass loss rate is aproximately two times higher than the optically thin estimate. The correlation thus holds between mass loss rates and $^{12}$CO line intensity ratio. Therefore, we introduce a new possible method for determining mass loss rates with short analysis time and low uncertainty.

\begin{figure}[h!]
 \centering
 \includegraphics[width=0.60\textwidth]{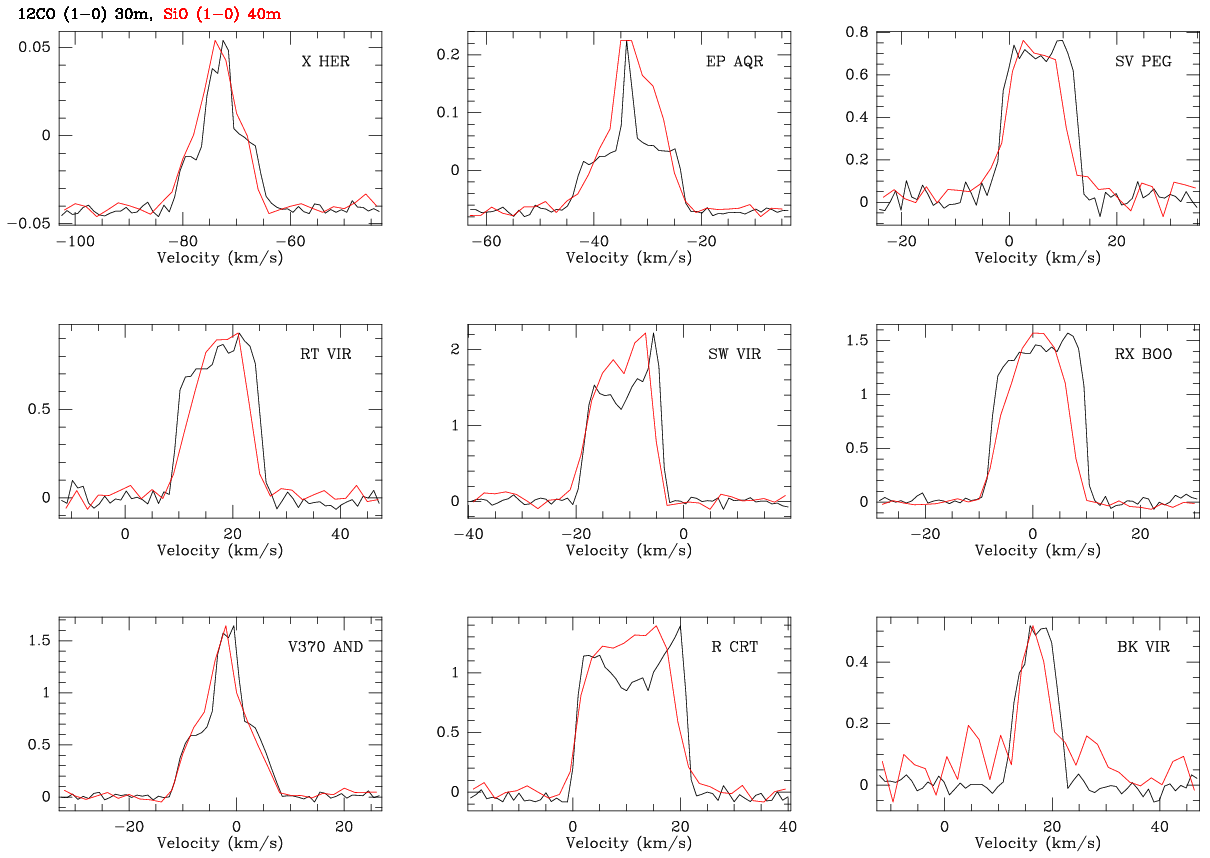} 
 \caption{IRAM 30m $^{12}$CO \textit{J}=1-0 (in black) and Yebes 40m SiO \textit{J}=1-0 (in red) spectra from the 9 sources where SiO is detected with high S/N ratio. $^{12}$CO units are in \textit{T$_{mb}$} scale while SiO spectra have been scaled to fit the box; note the similitude between both profiles in some sources in spite of total different chemical distribution of the two species.}
   \label{fig1}
\end{figure}

The $^{12}$CO(1-0)-to-$^{12}$CO(2-1) intensity ratio is, at the same time, correlated with the wind density ($\dot{M}$/v$_e$; r=0.63) and the photodissociation radius (R$_{CO}$; r=0.64) for O-rich SRs. These results are in good agreement with the correlation between $\dot{M}$ derived from CO observations and dust infrared properties reported by Loup et al. (1993). Concerning C-rich SRs, the correlation between $\dot{M}$ and the $^{12}$CO(1-0)-to-$^{12}$CO(2-1) intensity ratio, as well as dust infrared properties, is less clear. The observed behavior of $\dot{M}$ is probably explainable by the effects of saturation of \textit{T$_{mb}$}(1-0) for large mass loss rates and low kinetic temperature. There could also be possible variations of the gas-to-dust ratio or changes in $\dot{M}$.

We found that for several stars the SiO profiles are very similar to the CO profiles, an issue which is not viable unless these are non-standard envelopes (see Fig. 2).

\begin{figure}[h!]
 \centering
 \includegraphics[width=0.16\textwidth]{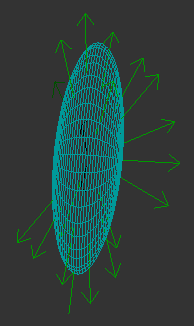} 
 \includegraphics[width=0.26\textwidth]{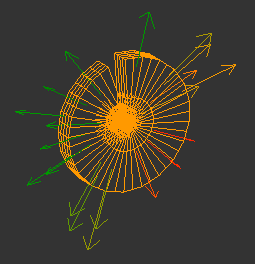} 
 \caption{Types 1 (AQ Sgr) and 3 ($\tau$$^{4}$ Ser) may be produced by spheroids and cylinders, respectively.}
   \label{fig1}
\end{figure}

According to {\tt{SHAPE}+\tt{SHAPEMOL}} (Steffen et al. 2011; Santander-Garc\'ia et al. 2015), types 1 and 2 profiles may be produced by thick and flattened spheroids, respectively. Type 3, however, may be produced by toruses and cylinders (see Fig. 3). Type 4 profiles may be produced by very complex structures. Types 3 and 4 may harbor binary systems ($\sim$41$\%$ of the sample).

\end{document}